\documentclass{amsproc}  %%% This is the only difference for the gr-qc version.

\copyrightinfo{2000}{John.W.Barrett}

\theoremstyle{definition}

\theoremstyle{remark}

\numberwithin{equation}{section}

\newcommand{\R}{{\mathbb R}}
\newcommand{\C}{{\mathbb C}}

\newcommand{\SO}{{\rm SO}}

\newcommand{\dd}{{\rm d}} % \d already defined

\input xy \xyoption{all} \xyoption{poly} % use Xy-pic
\input epsf
\newcommand{\epsfcenter}[1]{{\vcenter{\hbox{\epsfbox{#1}}}}} %center an epsf figure

\begin{document}

\title{State Sum Models and Quantum Gravity}

% Remove or comment out any unused author tags.
% author one information
\author{John W. Barrett}
\address{School of Mathematical Sciences
University Park
Nottingham NG7 2RD
UK}
%\curraddr{}
\email{jwb\@maths.nott.ac.uk}
%\thanks{}

% Use this \subjclass if you are using amsproc version 2.0 (December 1999).
%\subjclass[2000]{}
% Use this one if you are using an older version of amsproc.
\subjclass{}
\date{7 September 2000}

\begin{abstract}  This review gives a history of the construction of quantum field theory on four-dimensional spacetime using combinatorial techniques, and recent developments of the theory towards a combinatorial construction of quantum gravity. 
\end{abstract}

\maketitle

\section{State sum models}

In this short review I give a brief survey of the history of state sum  invariants of four-manifolds and the attempts to modify them to give models for quantum gravity. I emphasise at the outset that these are at present just models; we do not yet know how far they incorporate all the desirable features of a quantum theory of gravity. For brevity, the review will ignore the long and distinguished lower-dimensional history of these ideas. 

\subsection{States and weights} The general framework is as follows. Let $\sigma_n$ be a standard $n$-simplex, with vertices $0,1,2,\ldots,n$. The state sum model requires a set of \textit{states} $S$ to be given for each simplex. These states can be thought of either as the states of a system in statistical mechanics, or as a basis set of states in quantum mechanics.

This set of states is the same for any simplex of the same dimension, so one just has to specify the set of states $S(\sigma_n)$ for each $n$, up to the dimension of the space-time, $n=4$. The idea is that a state on a simplex specifies a state on any one of its faces uniquely; hence there are maps
$$\partial_i\colon S(\sigma_n)\to S(\sigma_{n-1}),$$
for each $i=0,\dots,n$, the $i$-th map corresponding to the $i$-th $(n-1)$-dimensional face (opposite the $i$-th vertex). These satisfy some obvious relations, and the whole setup is called a simplicial set.

\begin{figure}[ht]
$$\xymatrix{ 
& *{0} 
\ar@{-} 
[ddl]
_{\partial_2s} 
\ar@{-}
[ddr]
^{\partial_1s} 
\\
&*{s}
 \\
*{1}
\ar@{-}
[rr]
_{\partial_0s}
&&
*{2}
\\}
$$
\caption  {A state $s$ for a triangle} \label{atriangle}
\end{figure}
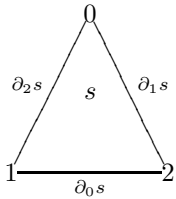

A \textit{weight} is a complex number which gives an amplitude (or Boltzmann weight) to a state.
$$w\colon S(\sigma_n)\to\C.$$

A state sum model uses the states and weights as the information for constructing a functional integral on a triangulated 4-manifold $M$. A \textit{configuration} on $M$ is an assignment of states to all the simplexes in the triangulation such that the states on the faces of any simplex are given by the boundary maps $\partial_i$. This means that the states on intersecting simplexes are related, because the common boundary data must match. 

The functional integral (or partition function) is a complex number calculated as follows
$$ Z(M)=\sum_s \prod_\sigma w\left(s(\sigma)\right).$$
The summation is over the set of all configurations $s$.  The product over simplexes $\sigma$ includes a weight factor for each simplex of every dimension.

\subsection{Topological models} If the manifold $M$ has a boundary, the states on the boundary are kept fixed, and the summation over states is only over the states for the interior simplexes. These states on the boundary are the boundary data for the quantum field theory. A model is said to be \textit{topological} if this partition function
$$Z\left(M,s(\partial M)\right)$$
depends only on the triangulation of the boundary, $\partial M$, i.e. not on the details of the triangulation of the interior.
The first idea for constructing such a state-sum model (in four dimensions) was the construction of Dijkgraaf-Witten for generic dimension, based on a finite gauge group\cite{DW}. This was generalised to the Lie group $SU(2)$ (where the model is unfortunately not finite) by Ooguri\cite{O}, and then to the $q$-deformed version, the Crane-Yetter state sum, which is again finite\cite{CY,CYK,R}. 

These models, based on a group or Hopf algebra, all have the feature that there is only one state for a 0- or a 1-simplex. The set of states states for a 2-simplex is the set of irreducible representations.
The set of states for a 3-simplex is a quadruple of representations, together with a basis element of the invariant elements in the tensor product of the four representations $h\colon\C\to a_1\otimes a_2\otimes a_3 \otimes a_4$.
$$S(\sigma_3)=\{(a_1,a_2,a_3,a_4,h)\}.$$

\begin{figure}[ht]
$$\epsfbox{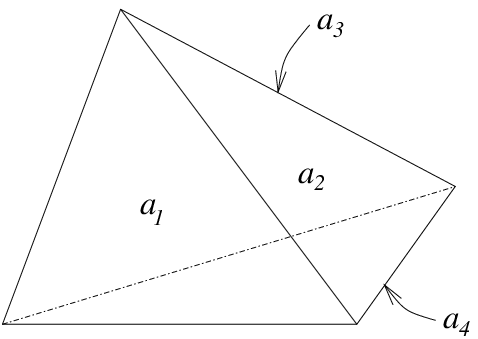}$$
\end{figure}

Finally, the set of states for the four-simplex has no new information, it is just the set of possible states for its boundary. The boundary maps are the obvious ones, for example
$$\partial_i(a_1,a_2,a_3,a_4,h)=a_i.$$ 
The most important weight is the weight for the 4-simplex, given by an inner product of the five states on the boundary tetrahedra. This inner product is best described by a diagrammatic calculus called \textit{spin networks}. The state on each tetrahedron is represented by the diagram
%\begin{figure}[ht]
$$\epsfbox{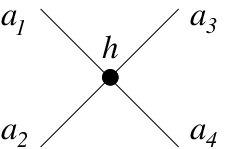}$$
%\end{figure}
called a \textit{vertex}. 
Each edge of the vertex corresponds to a triangle in the 4-simplex, and so the five vertices can be connected together by joining the edges pairwise. The amplitude is thus calculated by using the appropriate inner product pairwise on factors in the tensor product of all five vertices. However, as we have learnt from the study of quantum groups and knot theory, the order of the factors in the tensor product is important. The inner products can only be applied to adjacent factors, and when permuting the factors an $R$-matrix has to be used. All these features are captured in the spin network diagram, which is constructed by connecting the vertices together on $S^3$, and then projecting $S^3\to\R^2$ to give a generalised knot diagram. The amplitude can then be evaluated from the diagram using standard techniques developed from knot theory.

More recently, the Dijkgraaf-Witten model has been generalised to include non-trivial states on the 1-simplexes\cite{M}, and has a much more intricate structure. As yet, it is only known how to do this for finite groups. It will be interesting to see if this theory can be developed for Lie groups.

These models are closely related to some quantum field theories determined by a Lagrangian. The $BF$ Lagrangian is determined from a connection $A$ on a principal $G$-bundle over a 4-manifold $M$, and a 2-form $B$ of the same type as the curvature $F(A)$. It is
$$S=\int_M B\wedge F$$
in which the Lie algebra `indices' on $B$ and $F$ are contracted with an invariant metric on the Lie algebra.

The $BF$ functional integral $Z_{BF}$ is an average over the fields $A$ and $B$ of the phase factor $e^{iS}$, and the quantum field theory problem is to make sense of this average\cite{CCFM}. 

There is an intuitive picture of a correspondence of $BF$ theory with the state sum model based on the Lie group $G$. By geometric quantization, the irreducible representations of $G$ correspond to fluctuating vectors in the integral coadjoint orbits of the (dual) Lie algebra. In the state sum models, the irreducible representations are states for the triangles. In the $BF$ theory, one can integrate the $B$ field over a triangle to give precisely the same picture, namely a fluctuating element of the Lie algebra. With some more work, one can also see that both theories contain a connection\cite{O}. The $BF$ functional integral with a cosmological term is expected to be related to the Crane-Yetter state sum\cite{BZBF,FK}, however this correspondence is not understood yet in full.

\section{Quantum gravity} 

\subsection{Constrained state sum models} A tetrahedron embedded in $\R^4$ gives rise to edge vectors $X,Y,Z,\ldots$ as in the figure,
and hence to bivectors
$$p_1=*(X\wedge Y), \quad p_2=*(Y\wedge Z),\quad \ldots \qquad\in\Lambda^2(\R^4).
$$
\begin{figure}[ht]
$$\epsfbox{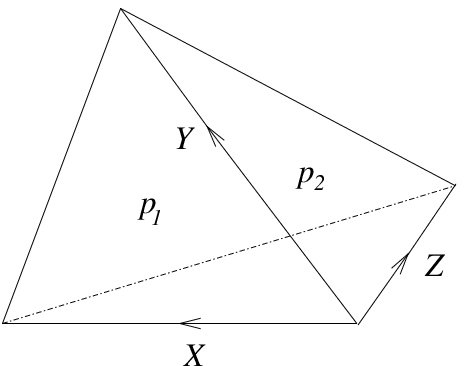}$$
\end{figure}
Bivectors are the same as anti-symmetric $4\times4$ matrices, which can be thought of as elements of the Lie algebra of $\SO(4)$. 

In quantum gravity, the geometry of a tetrahedon fluctuates, giving fluctuating bivectors, or fluctuating elements of the $\SO(4)$ Lie algebra, just as in the topological models. The difference is that here there are some constraints, such as $p_1\wedge p_1=0$ and $p_2\wedge p_1=0$. These constraints can be implemented on the set of states for the tetrahedron in the topological model\cite{BC,BZSF}, giving rise to the \textit{quantum tetrahedron}\cite{BAR}. The state sum that results from applying the constraints to the topological model is the model for quantum gravity, defined both for $\SO(4)$ (`Euclidean gravity') and
 $\SO(3,1)$ (`Lorentzian gravity')\cite{BC2}, using unitary irreducible representations. The $\SO(4)$ model also has a quantum group version, which one might expect to correspond to gravity with a cosmological term.
 
The Einstein-Hilbert Lagrangian can be written as a modification of the $BF$ Lagrangian with the group $SO(4)$ or $SO(3,1)$. Writing the elements of the Lie algebra as bivectors, the action has the local formula
$$ \int_M e^a \wedge e^b \wedge F^{cd} \; \epsilon_{abcd}$$
where $e$ is a frame field (vector-valued 1-form) on $M$. Thus if the $B$ field in the $BF$ theory is constrained by 
$$B^{ab}=(e^c\wedge e^d) {\epsilon^{ab}}_{cd}=*(e\wedge e)^{ab}$$
the $BF$ Lagrangian turns into the Einstein-Hilbert Lagrangian. These constraints are analogous to the constraints in the state sum model
\cite{RE2,DPF}.

The constraints are that the representations $a_i$ are \textit{simple representations}, those given by the vanishing of a certain Casimir operator, and $h=h_C$ is a certain \textit{canonical element}, which is unique\cite{RE}. The area of a triangle, given by $|p_i|$, is constant for an irreducible representation, and so corresponds to a second Casimir operator in the quantization. This means that the classical geometries in the state sum models are determined by the areas of triangles, not by the lengths of edges, (as in three dimensions\cite{PR}). The implications of this are deep; while the semiclassical limit for the 4-simplex weight gives exactly the expected Einstein action\cite{CYA,B,BWA,FK0}, the limit for a configuration on a triangulated 4-manifold gives a version of Regge calculus based on areas, whose geometric interpretation is not yet clear\cite{BRW}. Moreover, the quantum tetrahedron in four dimensions has just the four degrees of freedom, not the six one would expect from Euclidean geometry. The explanation of this fact is related to the uncertainty principle for four-dimensional geometries, and goes to the heart of the difficulty in providing a conventional geometrical explanation for the semiclassical limit\cite{BB}.  
A corollary of this is the non-existence of the Wheeler 3-geometry representation for the wave-function, a phenomenon first noticed in chiral quantum gravity\cite{ACZ}. 

The constraints can also be applied to spin networks in general: the \textit{relativistic spin network} evaluation is the invariant of graphs embedded in $S^3$ when each edge is labelled with a simple representation, and each vertex is the canonical $h_c$\cite{Y}.  In the case of the quantum group, the evaluation depends on the embedding, and is sufficiently powerful to distinguish pairs of embeddings of the same graph such as $\epsfcenter{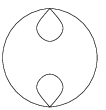}$ and $\epsfcenter{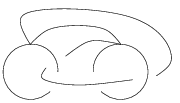}$, which do not contain any non-trivial knots or links\cite{BGR}.

\subsection{Matrix models and spin foams}  

The matrix model action 
 \begin{multline*}
 \frac12 \int \dd g_1\,\dd g_2\,\dd g_3\,\dd g_4\; \phi^2(g_1,g_2,g_3,g_4)\\
 +\frac\lambda{5!}\int\dd g_1\ldots\dd g_{10}\;\phi(g_1,g_2,g_3,g_4)\phi(g_4,g_5,g_6,g_7)\phi(g_7,g_3,g_8,g_9)\\
 \phi(g_9,g_6,g_2,g_{10})\phi(g_{10},g_8,g_5,g_1)
 \end{multline*}
 has the structure of a tetrahedron in its first `kinetic' term, and the structure of a 4-simplex in the second `potential' term. 
 This gives Feynman diagrams with topological structure a set of 4-simplexes which are glued together by identifying pairs of tetrahedral faces. These topological spaces are 4-manifolds with singularities on the lower-dimensional simplexes.

The momentum representation for Feynman diagrams amounts to a Fourier decomposition of the field $\phi$ into matrix elements for irreducible representations.  With suitable symmetries for the field $\phi$, the Fourier modes for individual Feynmann diagrams are exactly the configurations of Ooguri's topological state sum model with group $G$\cite{O}. The appearance of singular 4-manifolds is a natural generalisation for these models because there are no non-trivial states on the $0$- and $1$-simplexes, and so the manifold structure around these simplexes is not necessary. State sum configurations on these singular 4-manifolds can be described in an equivalent way as state sum configurations on the dual 2-complex\cite{DP}, which are examples of \textit{spin foams}\cite{BZSF}.  It is possible to modify the matrix model to impose the quantum gravity constraints, both for $G=\SO(4)$\cite{DFKR,PRO} and for $\SO(3,1)$\cite{PRO2}. In fact, the matrix model formulation is sufficiently flexible to allow a large class of simplex weights\cite{RR2}. Physical questions can be addressed by considering sums with fixed boundary data, which can be reformulated in terms of connections\cite{RR3}. 

\subsection{Triangulation independence}
A central question for non-topological state sum models or spin foam models is how to regain independence from the triangulation, as this is surely not physical. This problem does not yet have a convincing solution but there are several avenues open.
 
The matrix models are a presciption for summing over triangulations, including different topologies, and defining precise weighting factors for each configuration. This eliminates the dependence on one specific triangulation, but leaves the question of whether the partition function is convergent. 

There is a striking convergence with the ideas developed from the canonical quantization programme in the framework of chiral gravity, in which spin networks and spin foams play a role\cite{RR,RE0,D}. This leaves the possibility of a continuum reformulation of these ideas, perhaps by taking a `continuum limit'.
 
The remaining possibilty is that the constrained state sum model is actually just part of a topological state sum model, in which the remaining states describe the matter\cite{C}. Such an idea would bring us back to the topological and categorical roots of the subject.

\bibliographystyle{amsalpha}

\end{document}